# Common Investigation Process Model for Internet of Things Forensics

Muhammed Ahmed Saleh
Faculty of Engineering, School of Computing,
Universiti Teknologi Malaysia, Johor,
Malaysia
asmuhammed2@graduate.utm.my

Siti Hajar Othman
Faculty of Engineering, School of
Computing, Universiti Teknologi
Malaysia, Johor, Malaysia
hajar@utm.my

Arafat Al-Dhaqm
Faculty of Engineering, School of
Computing, Universiti Teknologi
Malaysia, Johor, Malaysia
mrarafat1@utm.my

Mahmoud Ahmad Al-Khasawneh
Faculty of Computer & Information
Technology
Al-Madinah International University
Shah Alam, Malaysia
mahmoud@outlook.my

*Abstract*— **Internet of Things Forensics (IoTFs) is a new discipline in digital forensics science used in the detection, acquisition, preservation, rebuilding, analyzing, and the presentation of evidence from IoT environments. IoTFs discipline still suffers from several issues and challenges that have in the recent past been documented. For example, heterogeneity of IoT infrastructures has mainly been a key challenge. The heterogeneity of the IoT infrastructures makes the IoTFs very complex, and ambiguous among various forensic domain. This paper aims to propose a common investigation processes for IoTFs using the metamodeling method called Common Investigation Process Model (CIPM) for IoTFs. The proposed CIPM consists of four common investigation processes: i) preparation process, ii) collection process, iii) analysis process and iv) final report process. The proposed CIPM can assist IoTFs users to facilitate, manage, and organize the investigation tasks.**

## I. INTRODUCTION

The IoT system is currently dynamically distributed across heterogeneous environments. As a result, an open environment and restricted resource makes using IoT vulnerable to attacks. Conducting digital investigations using existing tools and resources [1]–[3] has become difficult due to the dispersed and heterogeneous features of the IoT [3]–[5]. Law enforcement agencies and investigators face many challenges as a result of the existing IoT challenges [6]–[9]. The logic of connecting small devices to the Internet has open vistas of challenges raining from behavioural attributes [10]–[14] to technical components [5], [15]–[17]. IoT forensics (IoTFs) is a division of Digital Forensics (DFs) that investigates internet of things content to provide proof of internet crimes. It is deemed to be a significant area for identifying, acquiring, evaluating, and reconstructing internet of things events and exposing intruders' activities [18], within the broad scope of small scale device forensics [19], [20][21], owing to security attacks ranging from malicious software, botnet attacks [22][23][24]etc, which shows the need for conducting forensic readiness too. The IoTFs domain has been faced by several problems. There are numerous obstacles in the way of effective IoTFs, especially the lack of digital forensic resources that are well-suited to the heterogeneous and complex nature of the IoT environment [2], [5], [8], [16]. While the vast number of IoT devices available offers sufficient proof, it raises concerns about data management and detecting in a distributed environment, compromised devices. Several recent studies have suggested new investigative models or surveyed current problems in IoTFs to adapt digital forensics to the IoT system [25]–[27]. Several works have been developed for IoTFs field. For example [28] provided a series of IoT cybercrime scenarios that were carried out by a perpetrator who used different IoT to commit cybercrime. The authors used these scenarios to classify alternative sources of proof in the IoT system. The authors then used this data to develop a three-zone IoT investigation model, with first zone representing the internal network, second zone representing all hardware and software on the network edges, and the third zone represents hardware and software outside the internet. They stated that segmenting the attack area into First-Second-Third zones allows investigators to work more effectively and rapidly. Similarly, study in [16] suggested an IoT investigation system with a Digital Forensic Readiness (DFR) is a capability for planning and preparing for potential IoT cybercrime [29][30]. In addition, [31] suggested a real-time model for investigating IoT forensics. Their system was placed in place to keep track of the digital evidence collected during the investigation. Also, they talked about particularly during the pre-investigation, IoT forensic readiness. Also, using the ISO/IEC 27043 standard as a guide [32][33] suggested a holistic IoT device forensic model. Other similar studies which hinge on the ISO/IEC 27043 focusses on the readiness potentials of digital forensics [9], [15], [34]–[38]. The three key steps in their proposed model are forensic readiness, forensic investigation, and forensic initialization. They claimed that their model could be tweaked to work with a variety of IoT applications. It can be seen from the above that previous IoTFI research approaches mainly discussed the field of the IoTFI from 3 perspectives: technology, research processes, and the dimensions underlined by [3], [18], [39]–[43][44]. The IoTFI field lacks a structured and unified model in which the field experts can facilitate, manage, share, and reuse the IoTFI field knowledge [45], similar to other digital forensic subdomains as articulated in [46]–[51]. Therefore, this paper aims to propose a common investigation process model for IoTFs field using the metamodeling method.

This paper is structured as follows: The introduction of the IoTFs field offered in Section I, whereas the proposing common investigation processes model has been discussed in Section II, finally, the conclusion and future work of this paper has been introduced in Section III.

## II. COMMON INVESTIGATION PROCESS MODEL FOR THE INTERNET OF THINGS FORENSICS FIELD

This section proposes a common investigation process model for IoTFs field metamodeling approach [52][53] [38], [54], [55]:

1) Identify and select IoTF models
2) Gather investigation processes from selected models
3) Mapping gathered investigation processes
4) Propose common investigation processes
5) Validate and evaluate the completeness of the proposed common processes

### 1) *Identify and select IoTFs models*:

In this step, we identify and collect IoTFs models and frameworks based on selecting criteria adapted from [56][7], [38], [57], [58][33][59][60][61]. The output of this step is ten (10) models and frameworks as shown in Table I.

TABLE I. IDENTIFIED AND SELECTED IOTFS MODELS

| Year | Model | Extracted Investigation Process | Processes |
|------|-------|--------------------------------|-----------|
| 2013 | [28] | Preparation process, Acquisition process, Investigation process, Reporting and storage | 4 |
| 2016 | [62] | Proactive process, IOT forensic process, Reactive Process, Concurrent Process | 4 |
| 2017 | [63] | Identification and inspection, Time-based, thing forensic, NBT forensic investigation, Final report | 4 |
| 2017 | [64] | Collection, Examination, Analysis, Reporting | 4 |
| 2017 | [65] | Preparation, Context-based collection, Data analysis, and correlation, Information Sharing, Presentation, Review | 6 |
| 2018 | [66] | device monitoring manager module, forensic analyzer module, evidence recovery module, case reporting module, communication module storage module | 5 |
| 2018 | [67] | Identification on an evidence, Collection process, Examination process, Analysis part | 4 |
| 2019 | [68] | Collection, Extraction, Analysis, Visualization, Abstraction | 5 |
| 2020 | [69] | Identification, Transmission, IoT communication, Design stack | 4 |
| 2020 | [70] | Audit framework, Access log audit, Access control connection, Performance analysis, Analysis ratio, Analysis time, Event ratio | 7 |
| **Total Processes** | | **45 Investigation Processes** | |

### 2) *Gather investigation processes from selected models*:

In this step, we gather and extract investigation processes from selected models based on criteria adapted from [71], [72]. Each model has different investigation processes. For example model [28] has four investigation processes: preparation process, acquisition process, investigation process, and reporting and storage. [62] includes four investigation processes as shown in Figure 1.

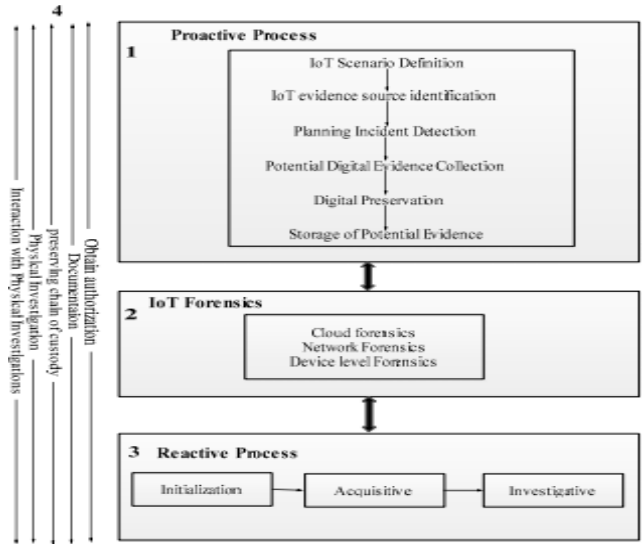

Fig. 1. Dimension of forensic investigation of the IoTs [62]

Also, [63] proposed a model which consists of four investigation processes as displayed in Figure 2: Identification and inspection Time-based thing forensic NBT forensic investigation Final report. Authors in [64] offered a model which consists of four investigation processes as shown in figure 3: collection, examination, analysis, reporting. Additionally, authors in [65] introduced a model which has 6 investigation processes: preparation, context-based collection, data analysis and correlation, information sharing, presentation, review. Authors in the model [67] proposed a model which has four (4) investigation processes: identification on evidence, collection process, examination process, and analysis part. Also, the authors in the model [68] proposed a model which consists of five (5) investigation processes as shown in Figure 4. Authors in models [69] and [70] proposed models with four (4) and seven(7) investigation processes respectively.

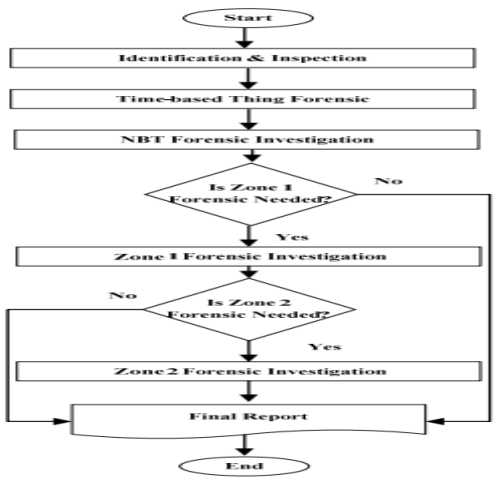

Fig. 2. Generic forensic framework for IoTs [63]

### 3) *Mapping gathered investigation processes*:

This step maps the extracted (45) investigation processes based on similarities and frequency [73], [74][10][38][75][76]. Investigation processes that have similar meaning/activities will map together and the highest investigation processes will propose as a common investigation process.

Table II displays the mapping process of the extracted processes. Four (4) investigation processes have the highest appearance amongst whole investigation processes which are: preparation process, collection, analysis, and final report. The preparation process appeared four times, the collection process appeared four times, the analysis process appeared 5 times, and finally, the final report process appeared three times. Next septs explain the initial proposing of common investigation processes for IoTFs domain.

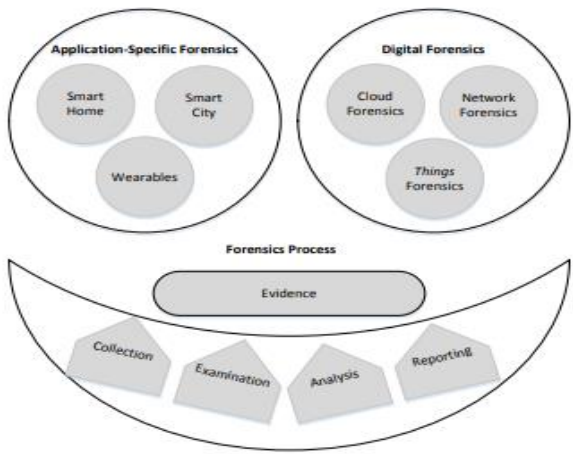

Fig. 3. Application-Specific Digital Forensics Investigative Model in the Internet of Things [64]

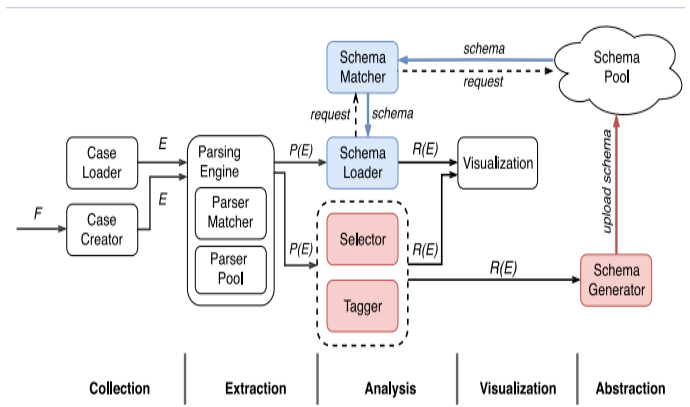

Fig. 4. Structure of the knowledge-sharing-based forensic analysis platform [68]

### 4) *Propose common investigation processes*:

The mapping processes performed in Step 3, highlighted four common investigation processes over 45 investigation processes as shown in Figure 5. The preparation process is used to prepare whole investigation resources, investigation team, trusted forensic toolkits, incident response plans, and seize investigation sources. The collection process is used to acquire and preserve whole seized data. The analysis process is utilized to reconstructing timeline events, analyze these events, and reveal who is the criminal. Finally, the whole investigation task will be summarized and concluded in the final report process.

### 5) *Validate and evaluate the completeness of the proposed common processes*:

The future work of this paper is to validate the completeness of the proposed common investigation processes. Approaches employed in [26] is a potential step towards achieving this step.

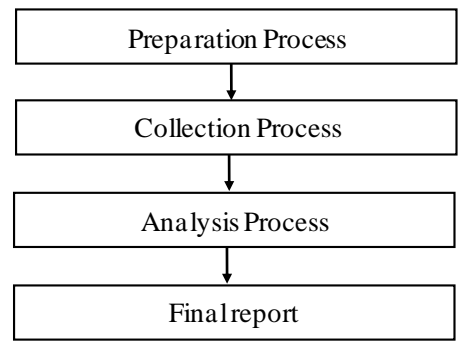

Fig. 5. Common investigation process model for IoTFs field

TABLE II. MAPPING PROCESS OF THE EXTRACTED INVESTIGATION PROCESSES

| Models/Process | [28] | [62] | [63] | [64] | [65] | [66] | [67] | [68] | [69] | [70] |
|---|---|---|---|---|---|---|---|---|---|---|
| Preparation process | √ | | | | √ | | √ | | √ | |
| Acquisition process | √ | | | | | | | | | |
| Investigation process | √ | | | | | | | | | |
| Reporting and storage | √ | | | | | | | | | |
| Proactive process | | √ | | | | | | | | |
| IoT forensic process | | √ | | | | | | | | |
| Reactive Process | | √ | | | | | | | | |
| Concurrent Process | | √ | | | | | | | | |
| Identification and inspection | | | √ | | | | | | | |
| Time-based thing forensic | | | √ | | | | | | | |
| NBT forensic investigation | | | √ | | | | | | | |
| Final report | | | √ | √ | √ | | | | | |
| Collection | | | | √ | √ | | √ | √ | | |
| Examination | | | | √ | | | √ | | | |
| Analysis | | | | √ | √ | | √ | √ | | √ |
| Information Sharing | | | | | √ | | | | | |
| Review | | | | | √ | | | | | |
| Device Monitoring Manager Module | | | | | | √ | | | | |
| Forensic Analyzer Module, | | | | | | √ | | | | |
| Evidence Recovery Module | | | | | | √ | | | | |
| Case Reporting Module | | | | | | √ | | | | |
| Communication Module | | | | | | √ | | | | |
| Storage Module | | | | | | √ | | | | |
| Visualization | | | | | | | | √ | | |
| Abstraction | | | | | | | | √ | | |
| Transmission | | | | | | | | | √ | |
| IoT communication | | | | | | | | | √ | |
| Design stack | | | | | | | | | √ | |
| Audit framework | | | | | | | | | | √ |
| Access log audit | | | | | | | | | | √ |
| Access control connection | | | | | | | | | | √ |
| Performance analysis | | | | | | | | | | √ |
| Event ratio | | | | | | | | | | √ |

## III. CONCLUSION

In this paper, we identified ten (10) IoTFs investigation process models. These models were identified and collected based on gathering criteria. The forty-five (45) common investigation processes have been extracted from the identified models. Then, four common investigation process model has been proposed based on mapping process. The proposed model consists of four investigation processes: preparation, collection, analysis, and final report. The future work of this paper is to validate the completeness of the proposed CIPM of the IoTFs field, as well as develop a structured and unified model called the Internet of Thinks Forensic Metamodel (IoTFM).